\documentclass[prd,aps,floats,epsfig,twocolumn]{revtex4}
\usepackage{graphicx}
\usepackage{color}
\usepackage{amsmath}
\newcommand{\beq}{\begin{equation}}
\newcommand{\eeq}{\end{equation}}
\newcommand{\bea}{\begin{eqnarray}}
\newcommand{\eea}{\end{eqnarray}}

\newcommand{\ApJ}{{\it Astrophys. J.\,}}

\newcommand{\PR}{{\it Phys. Rev.\,}}

\begin{document}
\title{Delayed Recombination and Standard Rulers}
\author{Francesco De Bernardis$^{1}$, Rachel Bean$^2$, Silvia Galli$^{1,5}$,
Alessandro Melchiorri$^1$, Joseph I. Silk$^3$ and Licia Verde$^4$}
\affiliation{$^1$ Universita' di Roma ``La Sapienza'', Ple Aldo Moro 2, 00185, Rome, Italy.\\
$^2$ Dept. of Astronomy, Space Sciences Building, Cornell University, Ithaca, NY 14853, USA.\\
$^3$ Astrophysics, Denys Wilkinson Building, University of Oxford, Keble Road, OX1 3RH, Oxford, UK.\\
$^4$ Institute of Space Sciences(IEEC-CSIC), Fac. Ciencies,
Campus UAB, Bellaterra, Spain\\
$^5$ Laboratoire Astroparticule et Cosmologie (APC), Universite' Paris Diderot - 75205 PARIS cedex 13.}

\begin{abstract}

Measurements of Baryonic Acoustic Oscillations in galaxy surveys have been recognized as
a powerful tool for constraining dark energy. However, this method relies
on the knowledge of the size of the acoustic horizon at recombination
derived from Cosmic Microwave Background Anisotropy measurements. This estimate is typically derived 
assuming a standard recombination scheme; additional radiation sources can delay recombination 
altering the cosmic ionization history and the cosmological inferences drawn from CMB and BAO data.
In this paper we quantify the effect of delayed recombination on the determination of dark energy parameters 
from future BAO surveys  such as BOSS and WFMOS. We find  the impact to be small but still not negligible. 
In particular, if  recombination is non-standard (to a level still allowed by CMB data), but this is ignored, 
future surveys may incorrectly  suggest the presence of a redshift dependent dark energy component.
On the other hand, in the case of delayed recombination, adding to the analysis one extra parameter 
describing deviations from standard recombination, does not significantly degrade the error-bars 
on dark energy parameters and yields unbiased estimates. This is due to the CMB-BAO complementarity.

\end{abstract}


\maketitle

\section{Introduction}

A key goal of modern cosmology is to investigate the nature of the
dark energy component, responsible for the current accelerated expansion
of the Universe. A variety of observables, from Cosmic Microwave Background
(CMB) anisotropies  \cite{wmap5komatsu,wmap5cosm}  to galaxy surveys
\cite{Tegmark:2003ud,Tegmark:2003uf,Cole:2005sx,Sanchez:2005pi,Tegmark:2006az},
will continue to be measured with increasingly refined
accuracy in the next years thanks to a coordinated effort  of satellite, ground based
and balloon-borne missions. Despite the fact that it accounts for about
$70 \%$ of the total energy density of the universe, dark energy is largely
unclustered and is typically measured just by its effect on the
evolution of the expansion history (i.e. the Hubble parameter) at redshift $z <2$. Since the cosmic
expansion depends on other key parameters as curvature or matter density, the
 nature of dark energy can therefore be revealed only by combination of
different observables and/or observations over a wide redshift range.

A recent development in the field of  large-scale galaxy surveys,  is to use the  measurement of 
Baryonic Acoustic Oscillations (BAO) \cite{Eisenstein:2005su,Percival:2006gs,Percival:2007yw} 
signal  at $z<2$ as a standard ruler. 
Oscillations in the primordial photon-baryon plasma leave an imprint
in the matter distribution. Since the frequency of these oscillations is related
to the size of the sound horizon at recombination, which is well constrained by CMB measurements, it
is possible to use the measurements of those oscillations at different redshifts
as a standard ruler and therefore determine the rate of the cosmic expansion.

It has been found that a number of theoretical systematics including non-linear growth, non-linear bias,
and non-linear redshift distortions can be efficiently modeled to minimize their contribution to uncertainties in the analysis of BAO observations \cite{Eisenstein:2006nk,Seo:2007ns,Seo:2008yx}. As such, they are one of the key observables of the next decade in cosmology and it is
therefore important to investigate how solid are the theoretical assumptions behind
this method. A crucial assumption, indeed, is the possibility of having a very accurate
meausurement of the size of the acoustic horizon at recombination. \cite{whitenstein} show that the interpretation of low redshift BAOs is robust  if CMB correctly determines  the  baryon-to-photon ratio and the matter-radiation equality but not the matter baryon or photon densities alone. CMB measurements of these quantities is generally robust, but , in view of high-precision data, it is important to quantify any possible systematic effects introduced by deviations from the standard evolution of the early universe . Previous papers have studied,
for example, the stability of the result under the hypothesis of an extra background
of relativistic particles or early dark energy (\cite{linder}).
In this case, the epoch of matter-photons equality is shifted and this may bias the
determination of size the acoustic horizon if unaccounted for.

In this paper we consider another possible mechanism that could, in principle, affect
the BAO  interpretation as standard rulers if not modeled, namely a possible delay,
or change in the recombination epoch that would result in a variation of the acoustic horizon.
Several papers, recently, have indeed considered this
hypothesis (see e.g. \cite{Seager00,Naselsky02,Dorosh02,bms,bms2,gbms,Zhang:2006fr}). Dark matter decay or annihilation,
black hole evaporation, or cosmic string decay can, for example, produce an
extra background of resonance photons that could delay recombination (see \cite{gbms} and references therein).
More exotic mechanisms as, for example, variations in the fine structure constant or in the
Newton constant, can also lead to a modified recombination epoch (\cite{alpha,zalzahn}).
We conclude that current constraints  on delayed recombination are compatible with a  change in the sound horizon which is  non-negligible when compared with the precision for future BAOs data.  This means that, if unaccounted for, delayed recombination could bias dark energy constraints from BAO.
On the other hand, if delayed recombination is accounted for in the joint CMB-BAO analysis form forthcoming surveys,  error-bars on dark energy parameters are only degraded by  less than $10$ \% compared to the standard analysis.

The paper is organized as follows: in the next section we discuss delayed recombination
and its effect on the sound horizon. In Sec. III we review the Fisher matrix formalim,
while in Sec. IV we forecast the impact of delayed recombination on dark energy parameters 
inference for several future
surveys. Finally, in Sec. IV, we draw together our findings and discuss the conclusions 
and their implications for future work.

\section{Delayed Recombination and the Sound Horizon}

In this section we modify the standard recombination picture by including the possibility of
 extra sources of resonance (Lyman-$\alpha$) photons. 
Note that we do not modify matter-radiation equality, an 
effect already explored by \cite{whitenstein}.

In the standard recombination scenario (see \cite{Peebles68,Zeldovich68}) 
the evolution of the electron ionization fraction, $x_{e}$ is given by:
\bea
-{dx_{e}\over dt}\left.\right|_{std}=C\left[a_c n x_{e}^{2}-b_c
(1-x_{e})\exp{\left(-{\Delta B\over k_{B}T}\right)}\right]
\label{eq1}
\eea
where $x_e$ is the electron fraction per hydrogen nuclei,
$n$ is the number density of atoms, $a_{c}$ and $b_{c}$ are
the effective recombination and photo-ionization rates for principle
quantum numbers $\ge 2$, $\Delta B$ is the difference in binding energy
between the $1^{st}$ and $2^{nd}$ energy levels and
\bea C={1+K\Lambda_{1s2s}n_{1s}\over1+K(\Lambda_{1s2s}+b_{c})n_{1s}}
\label{eqC},  \ \ \ \ K={\lambda_\alpha^{3} \over8\pi H(z)}
\eea
\noindent where  $\lambda_{\alpha}$ is the wavelength of the single Ly-$\alpha$
transition from the $2p$ level, $\Lambda_{1s2s}$ is
the decay rate of the metastable $2s$ level, $n_{1s}=n(1-x_{e})$ is
the number of neutral ground state $H$ atoms, and $H(z)$
is the Hubble expansion factor at a redshift $z$.

As in \cite{bms,bms2,gbms}, we include the possibility of extra resonance
(Ly-$\alpha$) photons at recombination with number density $n_{\alpha}$
which promote electrons to the $2p$ level (see \cite{Seager00,Naselsky02,Dorosh02,bms}):

\bea
{dn_{\alpha}\over dt}&=&\varepsilon_{\alpha}H(z)n
\label{eq0}
\eea

\noindent where $\varepsilon_{\alpha}$ is assumed constant with redshift.

This leads to a modified evolution of the ionization fraction:

\bea -{dx_{e}\over dt}&=&-{dx_{e}\over dt}\left.
\right|_{std}-(1-C)\varepsilon_{\alpha}H .
\ \ \ \ \ \ \ \label{eq2}
\eea

\noindent where ${std}$ refers to the the standard scenario.

The introduction of $\varepsilon_{\alpha}$ has the main effect of
delaying the redshift of recombination $z_*$. In the standard
scenario this redshift is given by (\cite{hu}):

\begin{eqnarray}
z_{*} & = & 1048(1+0.00124\omega_{\rm b}^{-0.738})(1+g_1\omega_{\rm m}^{g_2})\label{zrfit}\\
g_1 & = & 0.0783\omega_{\rm b}^{-0.238}/(1+39.5\omega_{\rm b}^{0.763})\nonumber\\
g_2 & = & 0.560/(1+21.1\omega_{\rm b}^{1.81})\nonumber
\end{eqnarray}

\noindent where $\omega_m$ and $\omega_b$ are the matter and baryon physical energy densities
respectively. Including positive values of $\varepsilon_{\alpha}$ decreases $z_*$ as (\cite{Seager00}):

\begin{equation}
z_*(\epsilon_\alpha)=(1+3\epsilon_\alpha)^{-0.042}z_*(\epsilon_\alpha=0)
\end{equation}

Conservative limits on $\epsilon_\alpha$ from current CMB data are of the order of
$\epsilon_\alpha < 0.5$ at $95 \%$ c.l. (\cite{gbms}). The redshift of recombination can therefore be smaller by $\sim 5 \%$ respect to the standard case. These constraints were obtained for a LCDM cosmology:  the bound on $\epsilon_{\alpha}$ would be relaxed even further for models where dark energy is not a cosmological constant.

\begin{figure}[hb]
\begin{center}
\includegraphics[width=9cm]{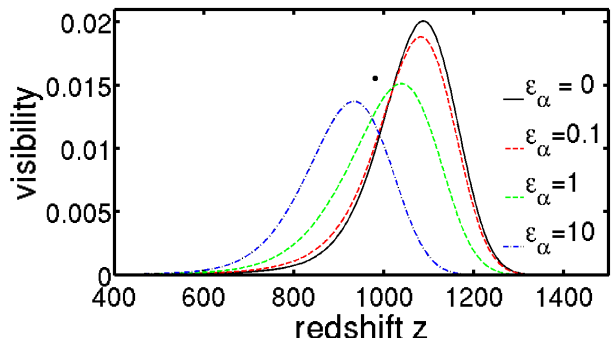}
\caption{Visibility function (defined as the density probability of photon last scattering), 
in function of $\epsilon_\alpha$. The inclusion of a delayed recombination
shifts the peak of the distribution at lower redshift and increases its width.}\label{vis}
\end{center}
\end{figure}

\begin{figure}[hb]
\begin{center}
\includegraphics[width=9cm]{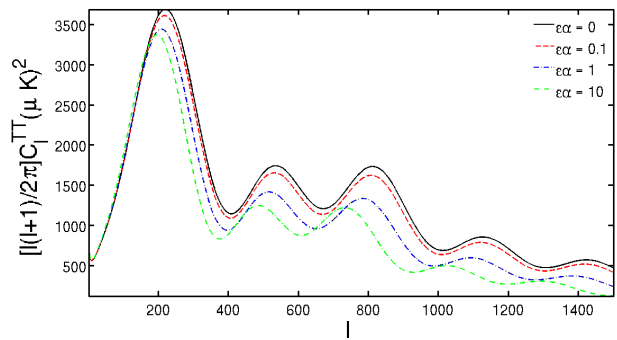}
\caption{Angular power spectrum of CMB anisotropies
in function of $\epsilon_\alpha$. The inclusion of a delayed recombination
shifts the peaks of the spectrum and damps the structure at small angular scales.}\label{cielle}
\end{center}
\end{figure}

We can easily understand the main effect of a delayed recombination by looking at
the variations in the visibility function which describes the density probability
for last scattering for a photon at redshift $z$. Its peak defines therefore the epoch 
of recombination. As we can see from Figure \ref{vis},
where we plot the visibility function defined as in \cite{selza} for different values
of $\epsilon_{\alpha}$, an increase in $\epsilon_\alpha$ shifts the visibility function
towards lower redshifts and increases the width of the distribution. We can therefore expect
two main effects on the CMB anisotropies: a displacement of the peaks
in the anisotropy and polarization power spectra due to the shift in the recombination epoch 
and a simultaneous damping of the anisotropies (see Fig. \ref{cielle}) 
due to the increase in the finite thickness of the last scattering surface.

In what follows we quantify the impact of this effect on the acoustic 
oscillations imprinted in the matter power spectrum that
can then be used as a ``standard ruler'' at lower redshifts.  
We investigate the variation of the sound horizon size $s$  as a 
function of $z_*$ and, therefore,  $\epsilon_{\alpha}$.
In fact galaxy surveys, by measuring
the scale along and across the line of sight, 
constrain $H(z)^{-1}/s$ and $D_A(z)/s$, where $D_A$ is the angular
diameter distance at redshift $z$. It is therefore straightforward to expect that
a systematic change in $s$ could bias the determination of $H(z)^{-1}$,
$D_A(z)$ and the derived parameters.

The size of the sound horizon $s$ at recombination 
can be approximated by the following formula (\cite{efstathiou}):

\begin{eqnarray}
s&=&\frac{c}{\sqrt{3}H_0}\Omega_{\rm m}^{-1/2}\int_0^{a_{*}}
\frac{da}{\sqrt{(a+a_{\rm eq}))(1+R(a))}} \nonumber \\
&\approx& \frac{19.8\;{\rm Mpc}}{\sqrt{\omega_{\rm b}\omega_{\rm m}}}\ln\left(\frac{\sqrt{R(a_{*})+R(a_{\rm eq})}
+\sqrt{1+R(a_{*})}}{1+\sqrt{R(a_{\rm eq})}}\right),
\label{rs}
\end{eqnarray}

\noindent where $H_0$ is the Hubble constant, $a$ is the scale factor
normalized to unity today, $R(a)=30496\omega_{\rm b} a$, the equality 
is at $a_{\rm eq}=1/(24185\omega_{\rm m})$ and $a_*=(1+z_*)^{-1}$.

\begin{figure}[hb]
\begin{center}
\includegraphics[width=9cm]{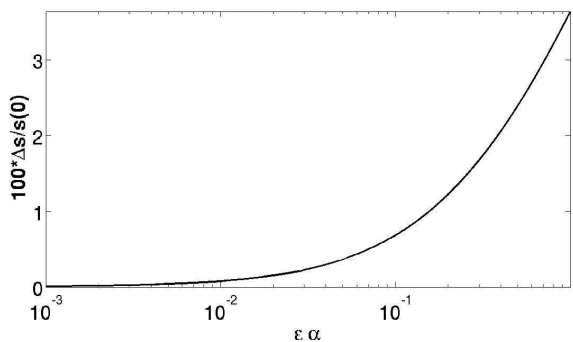}
\caption{Percentage of variation in the size of the acoustic horizon at recombination in function
of $\epsilon_\alpha$. Delayed recombination with $\epsilon_\alpha \sim 0.5$ could increase
the horizon by few percent.}\label{hor}
\end{center}
\end{figure}

Shifting the epoch of recombination, while leaving as unchanged the remaining parameters,
clearly changes the extrema of the integral and the size of the sound horizon.

In Figure \ref{hor} we show the percentage variation in $s$  as a function of $\epsilon_{\alpha}$: 
a value of $\epsilon_{\alpha}$ as large as $0.5$ yields a few percent variation in the 
CMB estimate of the sound horizon. This can potentially propagate into a large shift in the 
recovered dark energy parameters: at $z\sim 0.6$ a $\sim 1\%$ error in $s$ yields a 
$\sim 4\%$ error in $w$ assumed constant.  This shift is comparable to the  statistical 
error on this quantity expected from forthcoming surveys.
However, as we will see in the next sections and already from Figure \ref{cielle},
the impact of delayed recombination on observables is  broader. 
Indeed, the damping in the anisotropy power spectrum
introduces a degeneracy with other parameters as the scalar spectral
index $n_s$. As showed in \cite{gbms} considering values
of $\epsilon_\alpha \sim 0.1$ could mimic variations as large
as $\sim 10 \%$ in $n_s$ and in several other variables 
as the Hubble constant.

\section{Fisher Matrix Analysis for BAO and CMB: Standard Recombination}

In this section we briefly review the Fisher formalism mostly 
based on the seminal paper by \cite{Seo:2003pu} and
we refer the reader to this work for further clarifications.
At the end of this section we cross-check our results
with similar analyses already presented in the literature
and obtained under the assumption of standard recombination.

The usual definition of the Fisher matrix is (see \cite{fisher1,fisher2}):

\begin{equation}\label{fisher}
F_{\alpha\beta}\equiv\langle-\frac{\partial^{2}\ln L}{\partial p_{\alpha}\partial p_{\beta}}\rangle
\end{equation}

\noindent where $L$ is the likelihood function for a set of parameters $p_{i}$. If we suppose that the likelihood
has a maximum in some point of the parameter space $p_{i}^0$, the fiducial model, and if the data have a normal distribution
then the inverse of the Fisher matrix represents the covariance matrix of the parameters, i.e.:
\begin{equation}\label{cov}
\langle(p_{\alpha}-p^{0}_{\alpha})(p_{\beta}-p^{0}_{\beta})\rangle=(F^{-1})_ {\alpha\beta},
\end{equation}
as a consequence the statistical errors on the parameters are given by the square root of the diagonal elements of the inverse Fisher matrix. According to the Cramer-Rao inequality (\cite{cramer}) for unbiased estimators this is the best statistical error that one can obtain for the generic parameter $p_{\alpha}$:
\begin{equation}\label{sigma}
\sigma_{p_{\alpha}}\geq\sqrt{(F^{-1})_ {\alpha\alpha}}.
\end{equation}

For a galaxy survey the Fisher matrix can be approximated as (\cite{Tegmark:1997rp}):

\begin{equation}\label{fisher}
\begin{split}
F_{ij}=\int_{k_{min}}^{k_{max}}\frac{\partial \ln P(\vec {k})}{\partial p_i}\frac{\partial \ln P(\vec{k})}{\partial p_j}V_{eff}(\vec{k})\frac{d\vec{k}}{2(2\pi)^3}\\=\int_{-1}^{1}\int_{k_{min}}^{k_{max}}\frac{\partial \ln P(k,\mu)}{\partial p_i}\frac{\partial \ln P(k,\mu)}{\partial p_j}V_{eff}(k,\mu)\frac{k^2}{8\pi^2}dkd\mu
\end{split}
\end{equation}

\noindent where the derivatives of the linear matter power spectrum $P(k,\mu)$ are evaluated at $p_{i}^0$ and 
$V_{eff}$ is the effective volume of the survey, given by:

\begin{equation}
\begin{split}
V_{eff}(k,\mu)=\int\left[\frac{n_g(\vec{r})P(k,\mu)}{n_g(\vec{r})P(k,\mu)+1}\right]^2d\vec{r} \\=\left[\frac{n_gP(k,\mu)}{n_gP(k,\mu)+1}\right]^2V_{survey}
\end{split}
\end{equation}

\noindent where in the last equality is assumed that the comoving number density of galaxies $n_g$ is constant. The quantity $\mu$ is the cosine of the angle between the unit vector along the line of sight $\hat{r}$ and the wave vector $\vec{k}$, $\mu=\vec{k}\cdot\hat{r}/k$.
 The integration over $k$ that appears in (\ref{fisher}) is performed only up to a $k_{max}$, to exclude non-linear scales.
 The value of $k_{max}$ is redshift dependent and we calculate it using the same criterion of \cite{Seo:2003pu}, i.e. requiring $\sigma(R)=0.5$, where $R=\pi/2k$. The maximum scale of the survey, corresponding to $k_{min}$, doesn't influence the results and we can assume $k_{min}=0$.

We consider two future surveys: the Baryon Oscillation Spectroscopic Survey (BOSS \cite{BOSS}), that will cover redshifts up to $z=0.7$, and the Wide-Field Multi-Object Spectrograph (WFMOS \cite{Glazebrook:2005ui}) splitted in two samples, the first (WFMOS1) at redshifts $0.5<z<1.3$ and the second (WFMOS2) covering $2.3<z<3.3$. We therefore divide the redshift space in six bins, one up to $z=0.7$ for BOSS, four bins for WFMOS1 and one bin for WFMOS2. In Table $\ref{expspec}$ we list the survey area and the number of galaxies assumed in the analysis.

Following \cite{Seo:2003pu}, we consider $6$ cosmological parameters, the physical matter density $\Omega_mh^2$, the physical baryon density $\Omega_bh^2$, the matter fraction $\Omega_m$, the optical depth to reionization $\tau$, the scalar spectral index $n_s$ and the normalization $A_s$. The fiducial model is given by $\Omega_mh^2=0.1274$, $\Omega_bh^2=0.021$, $\Omega_m=0.26$, $\tau=0.05$, $n_s=0.96$, $A_s=2.4\cdot10^{-9}$ and $h=0.7$. We always assume flatness, i.e. $\Omega_{\Lambda}=1-\Omega_m$. In addition to these parameters we included other, redshift-dependent, parameters as the angular diameter distance $D_A(z)$, the Hubble parameter $H(z)$, the linear growth factor $G(z)$ and the linear redshift distortion $\beta(z)$. 
Our fiducial values for $H(z)$ and $D_A(z)$ are obtained assuming a cosmological constant model. For the bias we assume $b=2$ at $z=0.35$, $b=3$ for WFMOS2 at $z=2.8$ and values ranging from $b=1.2$ at $z=0.5$ to $b=1.7$ for $z=1.3$ for WFMOS1
With this parameterization each bin in redshift has a total of $10$ parameters, $6$ common to all bins and $4$ different for each bin.

The observed power spectrum is given by:

\begin{equation}\label{obsp}
P=\frac{D_{A,r}^2(z)H(z)}{D_A^2(z)H_r(z)}b^2(1+\beta\mu^2)^2\left(\frac{G(z)}{G(0)}\right)^2P^0(z=0,k)
\end{equation}

\noindent where the first factor accounts for the fact that the observed cosmology (subscript \textit{r}) could differ from the ``true'' cosmology, $b$ is the bias, given by $b=\Omega_m(z)^{0.6}/\beta(z)$, and the term $(1+\beta\mu^2)^2$ describes the linear redshift distortions. The linear growth factor $G(z)$ is given by the ratio $\delta(z)/\delta(0)$ between the linear density contrast at redshift $z$ and at $z=0$. We calculate the linear matter power spectrum at redshift $z=0$, $P^0(z=0,k)$, using the numerical code CAMB \cite{Lew}. For convenience we take the reference cosmology to be equal to our fiducial model.\newline
The resulting Fisher matrix is combined with the CMB information from the
Planck satellite \cite{:2006uk}. For a CMB experiment the Fisher matrix is given by \cite{fishcmb}:

\begin{equation}
     F^{CMB}_{\alpha\beta} = \sum_{l=2}^{l_{\rm max}} \sum_{i,j}
     \frac{\partial C_l^{i}}{\partial p_\alpha}
     ({\rm Cov}_l)_{i j}^{-1}
     \frac{\partial C_l^{j}}{\partial p_\beta},
\label{fishercmb}
\end{equation}
\noindent where the $C_l^{ij}$ are the well known power spectra for the temperature (TT), temperature-polarization (TE) and E-mode polarization (EE) ($i$ and $j$ run over TT, EE,TE) and ${\rm Cov}_l$ is the spectra covariance matrix.

\begin{table}
\begin{tabular}{lccccccclll}
  \hline \hline

Survey  & &   $z$   & &  $A(sq.deg.)$    &&   $N(10^6)$\\
  \hline \hline
BOSS    & &   $<0.7$  && $8000$                  &&     $1.5$     \\
WFMOS1    &&    $0.5<z<1.3$  && $2000$               &   &     $2.0$     \\
WFOMS2    &&    $2.3<z<3.3$  && $300$                &  &     $0.6$     \\

\hline
\end{tabular}\caption{Experimental specifications for the surveys used in this paper showing redshifts, survey area ($A$) and number of galaxies observed ($N$).} \label{expspec}
\end{table}

\begin{table}[htb]
\begin{center}
\begin{tabular}{rcccc}
& Chan. & FWHM & $\Delta T/T$ & $\Delta P/T$  \\
\hline
$f_{\rm sky}=0.65$
& 70 & $14'$ & 3.6 & 5.1 \\
& 100 & $9.5'$ & 2.5 & 4.0 \\
& 143 & $7.1'$ & 2.2 & 4.2 \\
& 217 & $5.0'$ & 4.8 & 9.8 \\
\hline
\end{tabular}
\caption{Experimental specifications for the Planck satellite. Channel frequency is given in GHz, FWHM in arcminutes, and noise in $10^{-6}$.}\label{Planck}
\end{center}
\end{table}

We use the  experimental configuration as described in Table \ref{Planck} with $l_{\rm max}=1500$ to calculate the sum in (\ref{fishercmb}).
For the CMB Fisher matrix we include as a free parameter also the angular diameter distance to the last scattering surface $D_{A,CMB}$.

\begin{table}[htb]
\begin{tabular}{lr}
Parameter&  $\sigma$\\
\hline
$\Omega_{m}h^2$&$0.0007$\\
$\Omega_{\Lambda}$&$0.013$\\
$w_0$&$0.11$\\
$w_1$&$0.14$\\
\end{tabular}
\caption{Uncertainties on dark energy and cosmological parameters from our Fisher matrix calculation in the case of standard recombination.}\label{sigma}
\end{table}

The full Fisher matrix is then simply given by the sum of the BAO Fisher matrices at each redshift bin and the CMB:

\begin{equation}
F_{ij}^{tot}=F_{ij}^{CMB}+\sum_{z=1}^NF_{ij}^z
\end{equation}

For $N$ bins the total number of parameters is then $6+N\times4+N_{CMB}$ where $N_{CMB}$ is the number of additional parameters for the CMB, in this case $D_{A,CMB}$ (and hence $N_{CMB}=1$). In the standard recombination case we have $6$ bins in redshift and our full Fisher matrix has $31$ parameters.\newline As explained in \cite{Seo:2003pu}, by taking $D_A(z)$, $H(z)$ and $G(z)$ as separate parameters at each redshift one is avoiding any assumption about a specific dark energy model. The main advantage of this procedure is in considering a Fisher matrix which is independent on the dark energy model and that can specified afterwards with a simple change of parameters. 
Here we allow for a possible time-dependent equation of state assuming a linear evolution with redshift:

\begin{equation}\label{stateq}
w(z)=w_0+w_1z
\end{equation}

To obtain constraints on dark energy we first select the set of parameters that affect distances and depend on dark energy, i.e. $\bar{p}=(\Omega_mh^2,\Omega_m,D_A(z)'s,H(z)'s)$ and we then marginalize over the remaining ``nuisance'' parameters. We implement this by inverting $F_{ij}^{tot}$ and selecting from the inverted matrix the elements that correspond to the parameters of interest. We then invert this submatrix to obtain a Fisher matrix ($F'_{nm}$) just for $\Omega_mh^2$,$\Omega_m$, $D_A(z)'s$ and $H(z)'s$. Finally we project in a new parameter space, for example $\bar{q} =(\Omega_mh^2,\Omega_\Lambda,wo,w_1)$, by equating the log likelihood functions in the old and in the new parameter space: $lnL(\bar{p})=lnL(\bar{q})$. In the Fisher matrix formalism this corresponds to a contraction of the old reduced Fisher matrix $F'_{nm}$ with derivatives between old parameters and new parameters:
\begin{equation}\label{fishernew}
F^{new}_{ij}=\frac{\partial{p_n}}{\partial{q_i}}F'_{nm}\frac{\partial{p_m}}{\partial{q_j}}
\end{equation}

The fiducial model is specified in the calculation of partial derivatives that appear in ($\ref{fishernew}$), we therefore assume a $\Lambda$-CDM model given by $w_0=-1$ and $w_1=0$. The results from our Fisher matrix analysis for the dark energy parameters are shown in Table \ref{sigma}. The results, obtained under the assumption of standard recombination, are substantially in agreement with those reported in \cite{Seo:2003pu}, with errors on $w_0$ and $w_1$ of a factor $\sim2$ smaller since we are considering improved experiments, with higher survey area and volumes. 
An improved Fisher matrix formalism has been recently presented in \cite{Seo:2007ns}, based on numerical simulations of \cite{Seo:2005ys}
and where nonlinear effects, bias and redshift distortions are more accurately implemented. The errors on $D_A$ and $H$ from their publicly available code for the 
BOSS survey at redshift $z=0.35$ are respectively $\sim1.1\%$ and $\sim2.1\%$ to be compared with ours $\sigma_D/D\sim 1.5\%$ and $\sigma_H/H\sim2\%$. We can therefore conclude that, for the scope of our paper, the results are in reasonable agreement.

\begin{table*}[ht]
\begin{center}
\begin{tabular}{c|ccccc||ccccccc}
\hline
$z$&&&Surveys only&               &          &&&    Surveys+CMB  &              &              \\
     &&$\sigma_D/D(\%)$ & $\sigma_H/H(\%)$ &  $\rho$ &&&  $\sigma_D/D(\%)$ & $\sigma_H/H(\%)$&  $\rho$ \\
\hline
     &&              &              &                &&&                    &              &              \\
$0.35$&& $8.39$           & $8.48$       &   $-0.964$      &&&       $1.48$        &  $1.95$              & $-0.151$     \\
$0.6$&& $8.57$           & $8.82$       &   $-0.890$      &&&       $2.76$        &  $3.48$            &   $-0.160$      \\
$0.8$&& $8.16$           & $8.27$       &   $-0.942$      &&&       $1.91$        &  $2.38$            &   $-0.157$      \\
$1.0$&& $8.10$           & $8.17$       &   $-0.957$      &&&       $1.70$        &  $2.07$            &   $-0.212$   \\
$1.2$&& $8.01$           & $8.06$      &   $-0.964$      &&&       $1.57$        &  $1.89$            &   $-0.225$  \\
$2.8$&& $7.93$           & $7.24$     &   $-0.979$      &&&       $1.16$        &  $1.29$            &   $-0.217$      \\
\end{tabular}
\caption{Relative uncertainties on $D_A(z)'s$ and $H(z)'s$ and their correlation coefficient ($\rho$) at each redshift. The correlation is large unless considering CMB observations, needed to calibrate the sound horizon.}\label{dah}
\end{center}
\end{table*}

Let us now describe in more detail the correlation between the different parameters in order to emphatize the
importance of the CMB measurement of the sound horizon.
In the BAO only Fisher matrix, strong correlations are present between the angular diameter distances $D_A(z)'s$ and $H(z)$'s due to the poor
determination of $\Omega_mh^2$, $\Omega_bh^2$ and $n_s$. These degeneracies  can be reduced by an accurate CMB measurement of these cosmological parameters. As we can see from Table \ref{dah} the correlations between the $D_A(z)'s$ and $H(z)'s$ (calculated as $\rho=(F^{-1})_{ij}/\sqrt{(F^{-1})_{ii}(F^{-1})_{jj}}$) from galaxy surveys alone are large, showing that, until the scale of sound horizon and other parameters are unknown, only the product $H(z)D_A(z)$ can be precisely determined from BAO surveys.
Is clear from the above discussion that if the CMB information is biased by the assumption of a incorrect model of recombination, it
will have an impact on the determination of the cosmic parameters from BAO. We investigate this in detail in the next section.

\section{Delayed recombination and cosmological inferences drawn from CMB and BAO data}


As shown in \cite{Heavens:2007ka} it is easy to compute the shift in the best fit parameters when other parameters
are fixed at a \textit{wrong} fiducial value within the Fisher matrix formalism. In general if we fix a number $p$ of parameters 
(say $\psi_{\gamma}$, with $\gamma=1,...,p$) to an incorrect value which differs from the true value for an amount
$\delta\psi_{\gamma}$, then the others $n$ parameters ($\theta_{\alpha}$, with $\alpha=1,...,n$) will be shifted of an amount given by:
\begin{equation}\label{shift}
\delta\theta_{\alpha}=-(F'^{-1})_{\alpha\beta}S_{\beta\gamma}\delta\psi_{\gamma}\hspace{10pt}\alpha,\beta=1,..,n
\end{equation}
\noindent where $(F'^{-1})_{\alpha\beta}$ is the inverse Fisher
matrix in the space of $\theta_{\alpha}$ parameters (an $n\times n$
matrix) and $S_{\beta\gamma}$ is a subset of the full $(n+p)\times
(n+p)$ Fisher matrix including also the $\psi_{\alpha}$ parameters.
Equation ($\ref{shift}$) shows that the shift on $\theta_{\alpha}$
parameters depends on the correlations among various parameters, on
how well these parameters are constrained and on how strong is the
dependence of the observable from the $p$ parameters fixed, i.e. on
the derivatives of the observable with respect to the
$\psi_{\gamma}$ parameters, which are contained in the
$S_{\beta\gamma}$ matrix. In our case we want to see what is the
effect of assuming standard recombination, i.e. fixing
$\epsilon_{\alpha}=0$, while its true value is different
from zero and we take it to be $\epsilon_\alpha=0.05$.

To study the
effects of a modified recombination on BAOs survey we follow the
method described above computing first a Fisher matrix for the parameters
listed in the previous section and then repeating the computation
including also $\epsilon_{\alpha}$. We then select from this second
Fisher matrix the sub-matrix $S_{\beta\gamma}$ that appear in
($\ref{shift}$), i.e. the row and the column corresponding to
$\epsilon_{\alpha}$, and calculate the shift for the distances,
Hubble parameters and dark energy parameters.\begin{table}[h]
  \begin{tabular}{lrr}
              $z$ &  $\delta D/D$  & $\delta H/H$ \\
\hline \hline
              $0.35$&   $-3.0\%$  & $3.0\%$ \\
              $0.6$ &   $-3.4\%$  & $3.3\%$ \\
              $0.8$ &   $-3.6\%$  & $3.6\%$ \\
              $1.0$ &   $-3.1\%$  & $3.0\%$ \\
              $1.2$ &   $-3.4\%$  & $3.3\%$ \\
              $2.8$ &  $-4.5\%$  & $4.5\%$ \\
              \hline
            \end{tabular}\caption{relative shift in the measure of $D_A(z)$ and $H(z)$ for the mean redhsift of each bin as computed from ($\ref{shift}$) assuming $\epsilon_{\alpha}=0$
            in the hypothesis that true value is $\epsilon_{\alpha}=0.05$.} \label{relshift}
\end{table}

\begin{figure*}[h]
  \begin{center}
    \begin{tabular}{lr}
      \resizebox{70mm}{!}{\includegraphics[scale=.3]{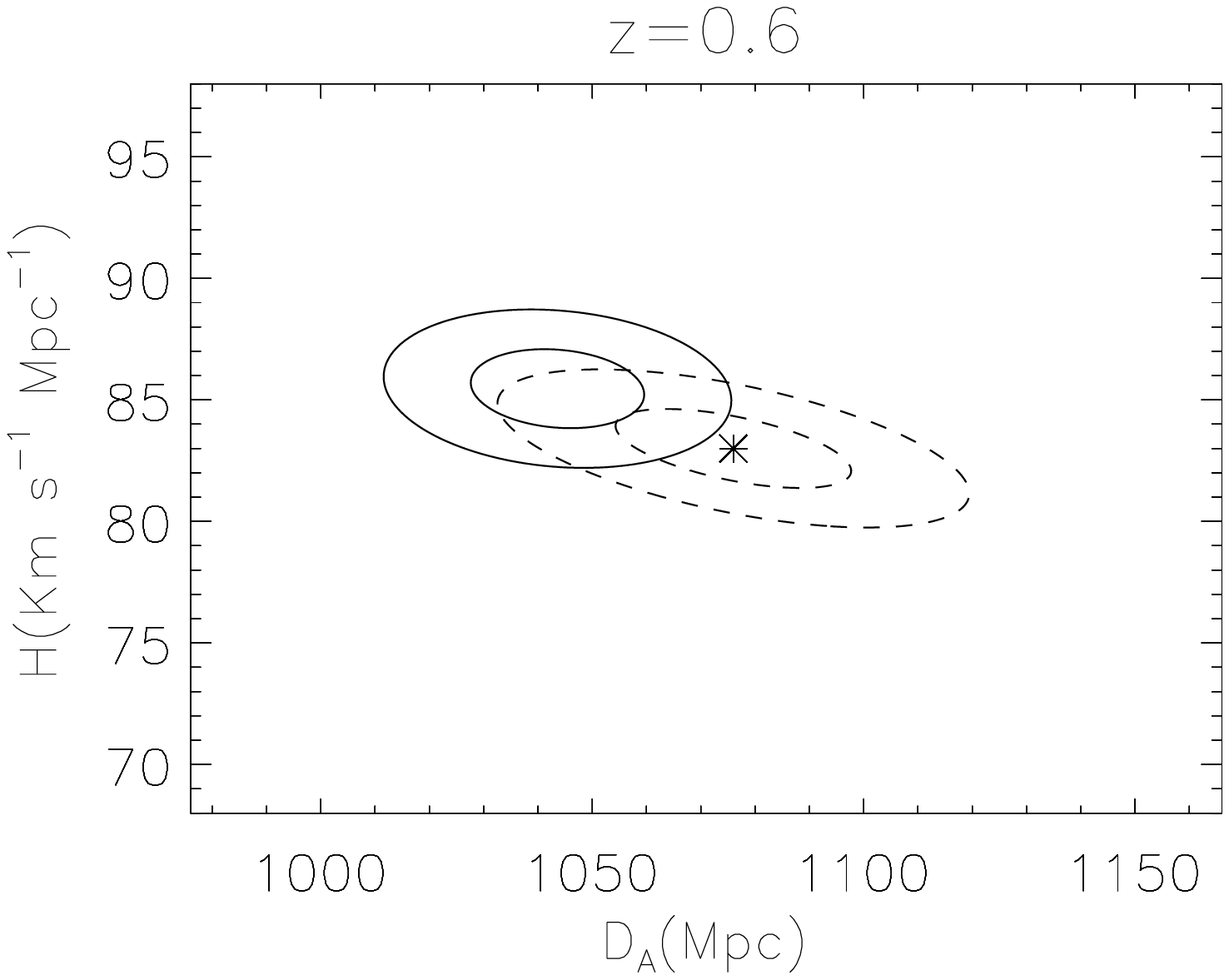}} &
      \resizebox{70mm}{!}{\includegraphics[scale=.3]{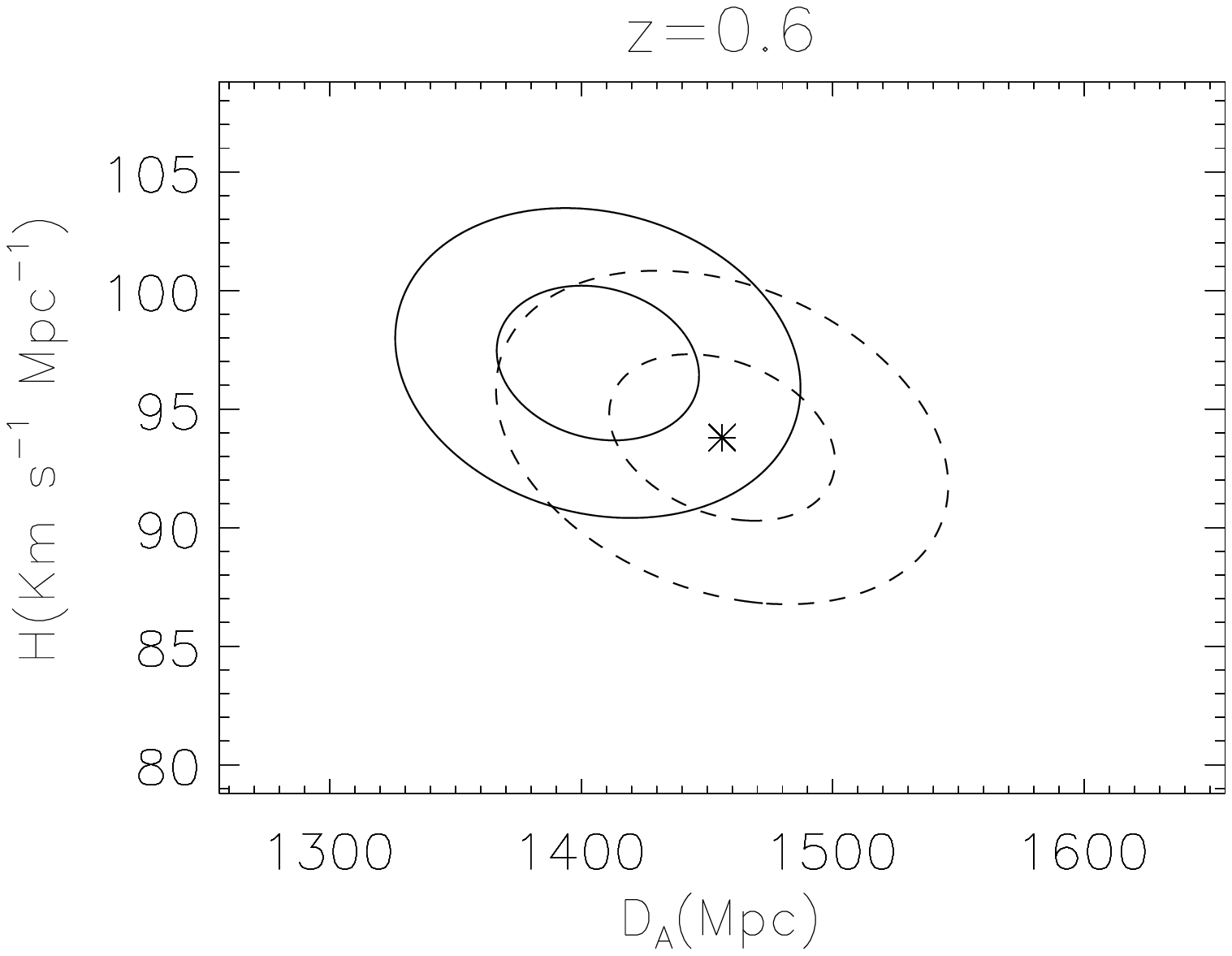}} \\
      \resizebox{70mm}{!}{\includegraphics[scale=.3]{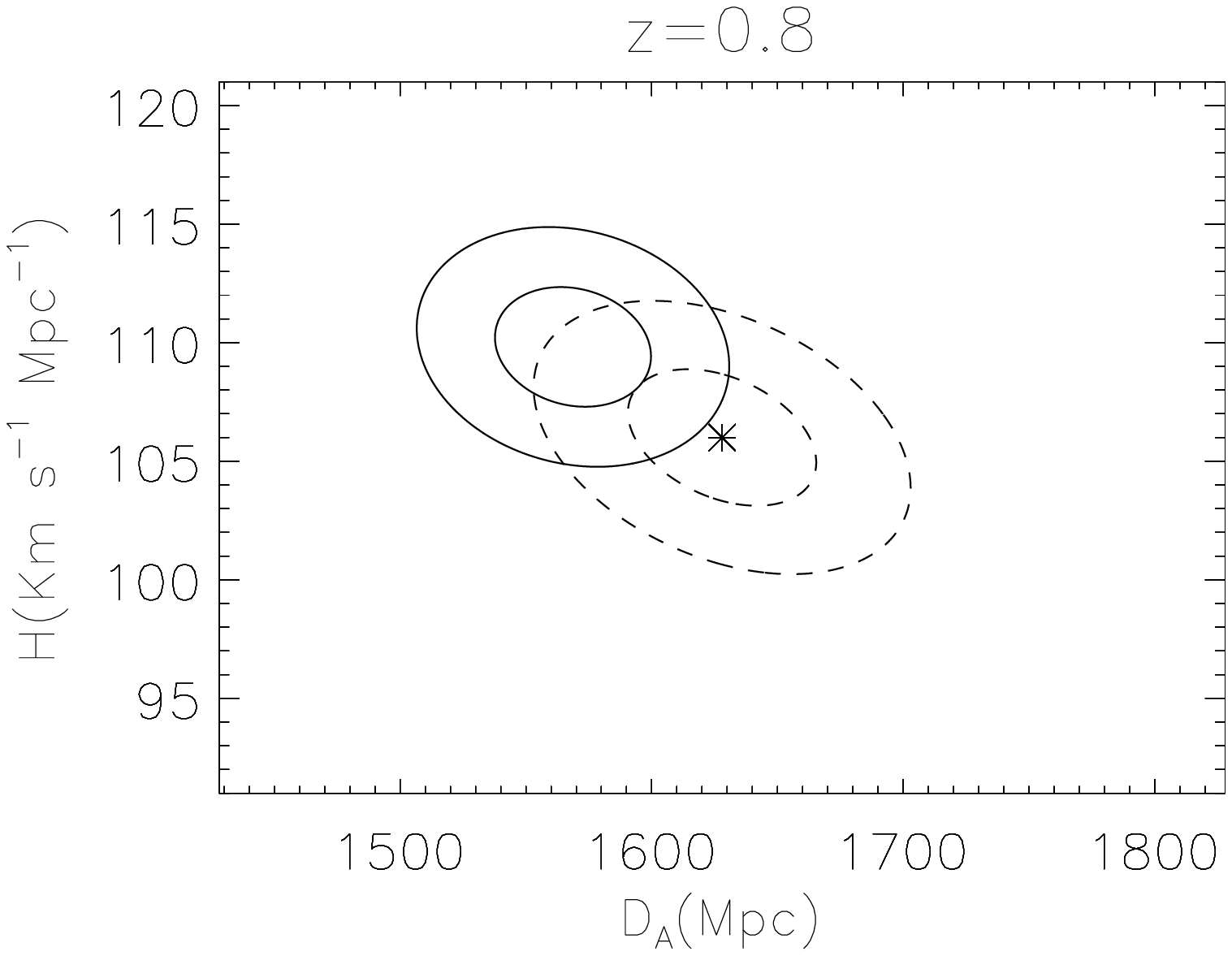}}  &
      \resizebox{70mm}{!}{\includegraphics[scale=.3]{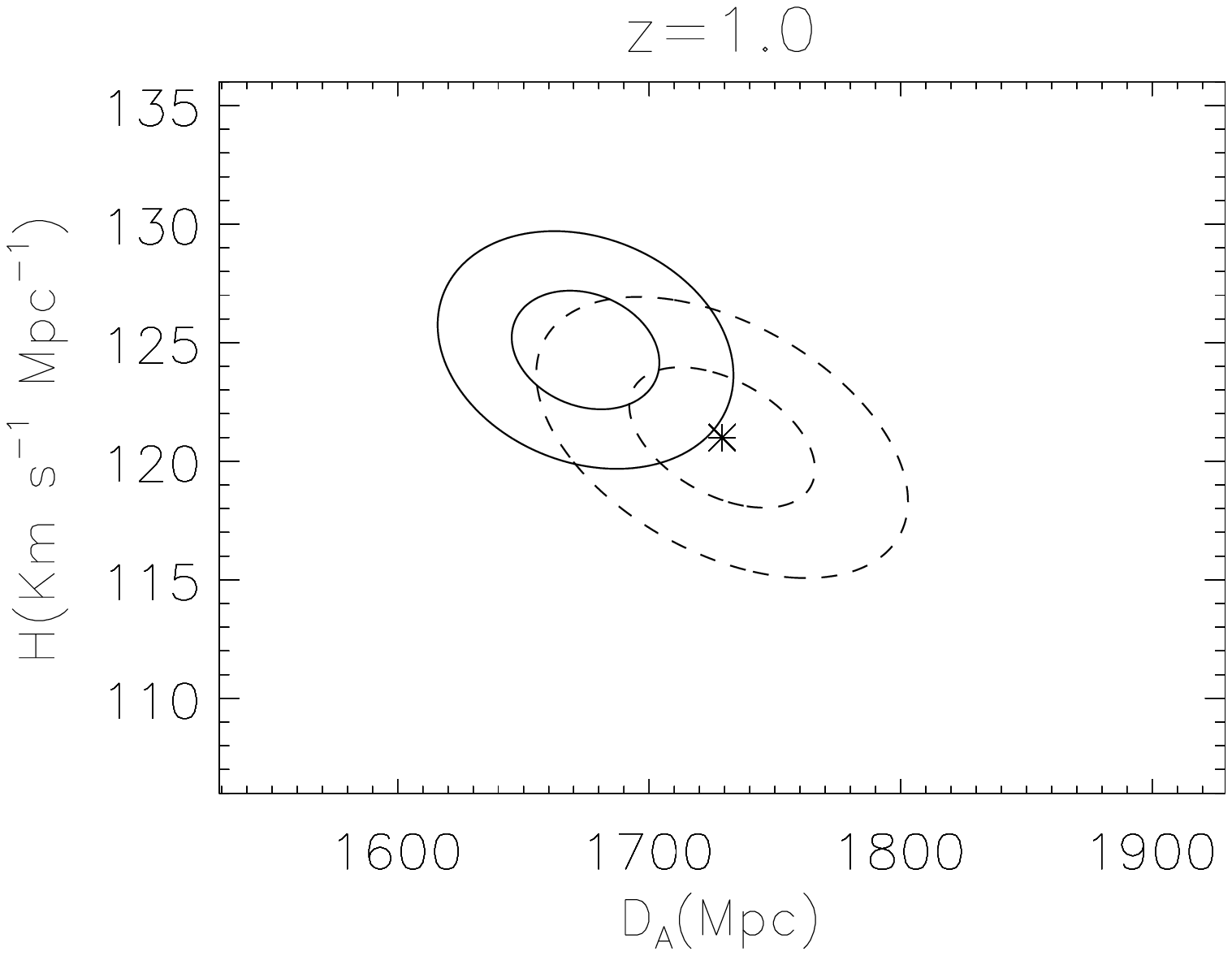}} \\
      \resizebox{70mm}{!}{\includegraphics[scale=.3]{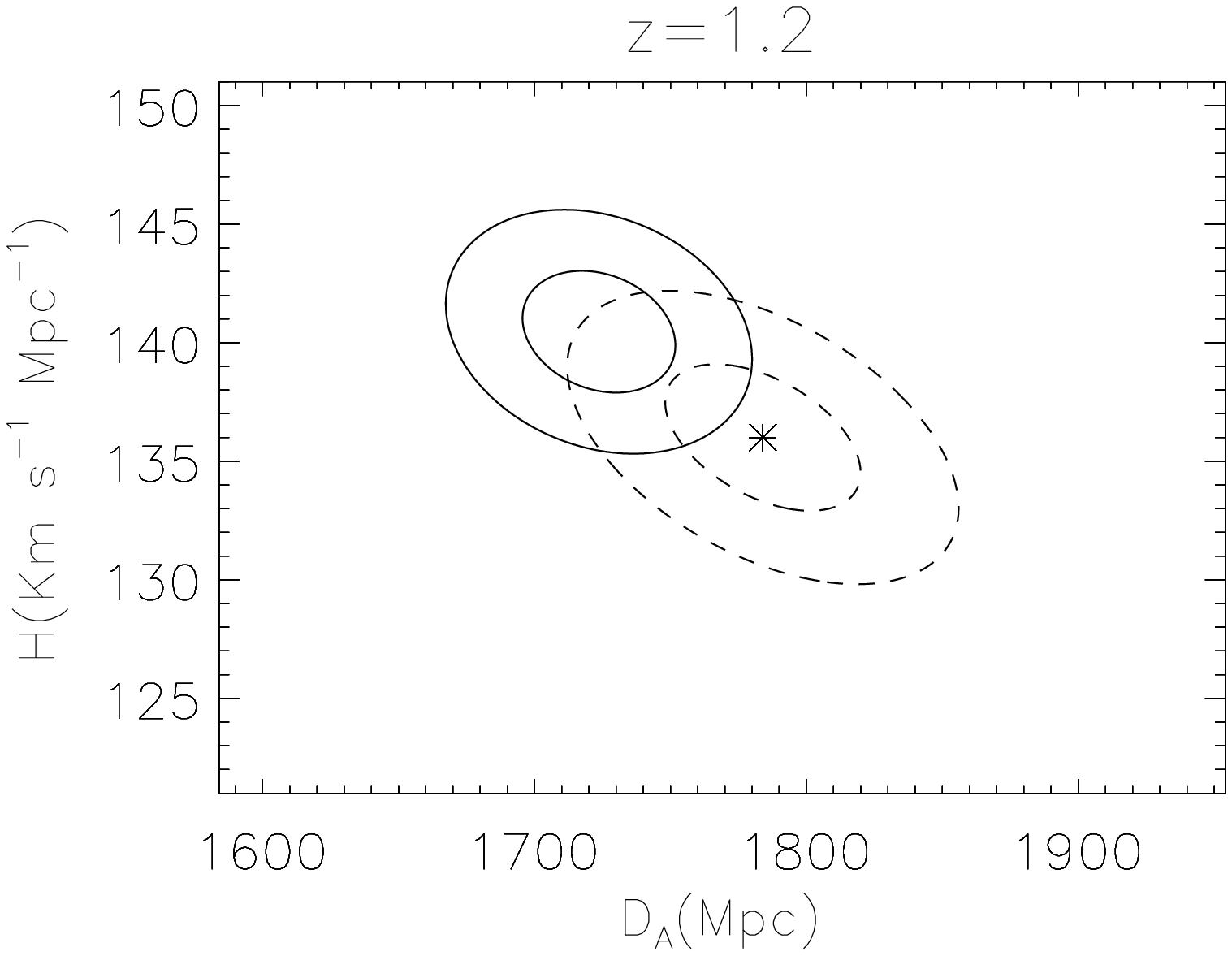}} &
      \resizebox{70mm}{!}{\includegraphics[scale=.3]{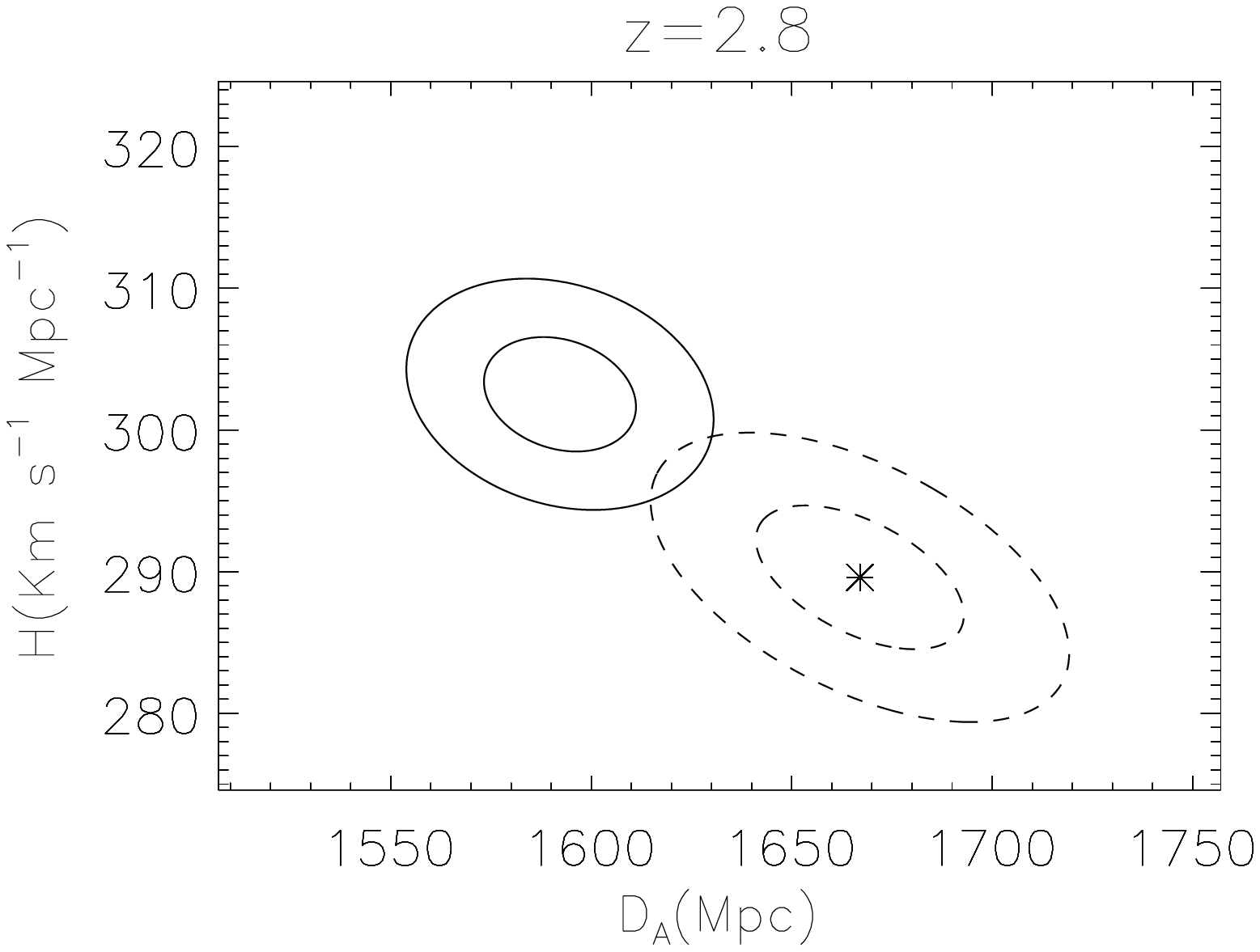}} \\
    \end{tabular}
    \caption{Shifts on best fit of $D_A(z)$ and $H(z)$ (solid line) for the mean redshift of each bin caused by a wrong
    determination of sound horizon when fixing $\epsilon_{\alpha}=0$ in the hypothesis that
true value is $\epsilon_{\alpha}=0.05$. Dashed contours are the
1,2 $\sigma$ constraints obtained assuming the correct recombination model while solid contours assume standard recombination, The
star is the correct fiducial model for $D_A(z)$ and
$H(z)$.}\label{plotshift}
  \end{center}
\end{figure*}

\begin{figure}[h]
\begin{center}
\includegraphics[width=9cm]{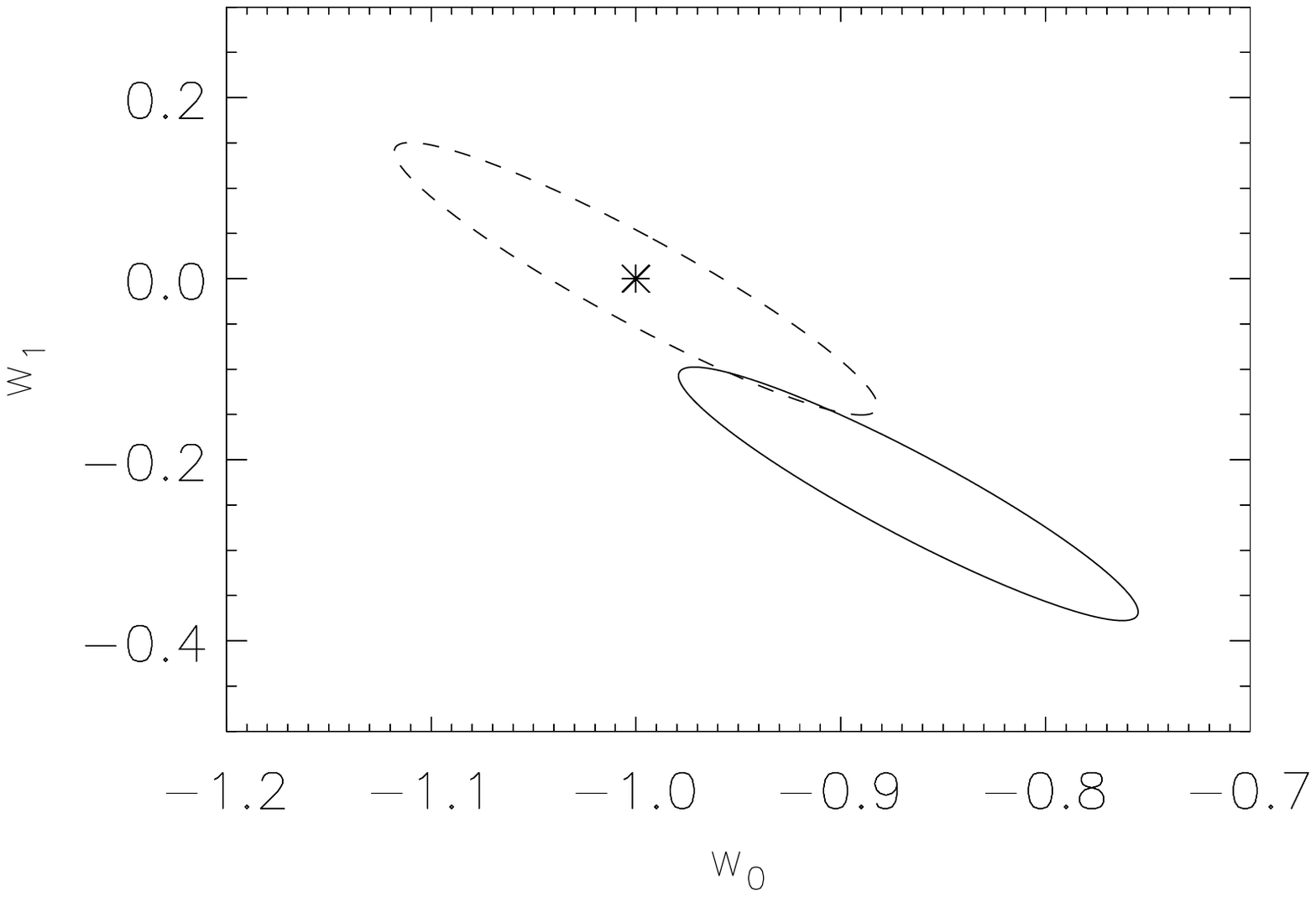}
\caption{Constraints at $68 \%$ c.l. on $w_0$ and $w_1$, assuming an incorrect
recombination model (solid line) and with the correct model (dashed line). The true model is indicated with the star.}\label{wow1}
\end{center}
\end{figure}
\newpage

We expect a shift $\delta s$ in the value of sound horizon to cause a shift in angular diameter distances and Hubble parameters, being
$D_A\propto s$ and $H\propto 1/s$. Moreover is easy to verify that
relative shift in $D_A(z)$ and $H(z)$ at each redshift should be of
the same extent and opposite: where $\delta D_A/D_A=-\delta H/H$. This is shown 
in Table V. Applying
($\ref{shift}$) to calculate shift in $D_A(z)$ and $H(z)$ confirms
this, as shown in Table \ref{relshift}. We show also absolute shifts
in $D_A$ and $H$ for each bin in Figure \ref{plotshift}. 
The shifts in $D_A$ and $H$ are at the $3\%$-$4\%$ level and, except for the last bin, are below the $95\%$C.L.
\newline
We remind that here we choose a fiducial model for the dark energy parameters with $w_0=-1$ and $w_1=0$.
Figure \ref{wow1} shows that assuming an incorrect recombination model introduces a shift in the recovered
value of $w_0$ towards $w_0>-1$ and of $w_1$ towards negative values, as a consequence of the shifts on $D_A$ and $H$. For $\Delta \epsilon_{\alpha}=0.05$ the effect is comparable to the  $1\sigma$ error on  $w_0$,  and slightly larger for  $w_1$. 
With the incorrect recombination model, the best fit are shifted to $w_0=-0.88$ and $w_1=-0.22$ respect to the
fiducial $\Lambda$-CDM model. An incorrect assumption on recombination could therefore mislead us towards claiming 
deviations of dark energy from a cosmological constant.

Even if this misinterpretation of the early universe physics induces  shifts $<95\%$C.L. both for $w_0$ and $w_1$, these results show that an incorrect calibration of sound horizon due to a wrong assumption for the recombination model could bias estimation of dark energy parameters by 10\% or more.

As shown in Figure \ref{wow1}, to recover an unbiased  best fit value  for dark energy parameters (in our case the fiducial model, $w_0=-1$ and $w_1=0$) the correct recombination model must be used and hence one has to  introduce an additional parameter, $\epsilon_{\alpha}$ in the analysis. Introducing this extra  degree of freedom could in principle lead to an increase in the uncertainties on others parameters.
We have verified that the increased uncertainty in the dark energy parameters is small.
 Including $\epsilon_\alpha$ in the analysis increase $1\sigma$
errors on these parameters less than  a $10\%$ percent yielding $\sigma_{w_0}=0.12$ and $\sigma_{w_1}=0.15$ (compare with \ref{sigma}). Forthcoming data sets will thus disentangle effects of non-standard recombination from the effects of dark energy.

\section{Conclusions}

In this paper we have investigated how a biased determination of the sound horizon due to a incorrect recombination model affects cosmological parameters measurements  from future BAO data. We have shown that assuming a standard recombination model when the true model is different can bias the measure of  the CMB sound horizon and the best fit value of cosmological parameters such as the Hubble parameter. This shift propagates in  a shift on the values of $D_A(z)$ and $H(z)$ and hence on the best fit values of $w_0$ and $w_1$ determined from BAO  surveys. We have shown that a deviation from standard recombination which is still allowed by current CMB data, propagates in a bias in $w_0$ and $w_1$  comparable  to  their $1\sigma$ statistical error from forthcoming surveys. An incorrect calibration of sound horizon due to a wrong assumption for the recombination model could bias estimation of dark energy parameters by 10\% or more. In this case  future surveys may thus  incorrectly  suggest the presence of a redshift dependent dark energy component.

To recover an unbiased  best fit value  for dark energy parameters the correct recombination model must be used or must be described by the set of cosmological parameters used in the analysis. 
We have concentrated on the case of delayed recombination where extra sources of resonance (Lyman-${\alpha}$) photons modify the evolution of the ionization fraction. This is well modeled by the addition of a single extra parameter. We have employed a Fisher matrix formalism to forecast errors and parameters biases for forthcoming CMB (Planck) and BAO (BOSS, WFMOS) surveys.
For this data set combination, introducing the additional  degree of freedom describing deviations from standard recombination does not significantly degrade the error-bars on dark energy parameters and yields unbiased estimates. 

This is due to the CMB-BAO complementarity: forthcoming data sets will thus disentangle effects of non-standard recombination from the effects of dark energy.

\section{Acknowledgements}

This research has been supported by ASI contract I/016/07/0 "COFIS".
RB's work is supported by NASA ATP grant NNX08AH27G, NSF grants AST-0607018 
and PHY-0555216 and Research Corporation. LV  acknowledges the  support of FP7-PEOPLE-2007- 
4-3-IRG n 20218 and of  CSIC I3 $\#200750I034$.

\end{document}